\def\year{2022}\relax
\documentclass[letterpaper]{article} % DO NOT CHANGE THIS
\usepackage{aaai22}  % DO NOT CHANGE THIS
\usepackage{times}  % DO NOT CHANGE THIS
\usepackage{helvet}  % DO NOT CHANGE THIS
\usepackage{courier}  % DO NOT CHANGE THIS
\usepackage[hyphens]{url}  % DO NOT CHANGE THIS
\usepackage{graphicx} % DO NOT CHANGE ETHIS
\usepackage{natbib}  % DO NOT CHANGE THIS AND DO NOT ADD ANY OPTIONS TO IT
\usepackage{caption} % DO NOT CHANGE THIS AND DO NOT ADD ANY OPTIONS TO IT
\DeclareCaptionStyle{ruled}{labelfont=normalfont,labelsep=colon,strut=off} % DO NOT CHANGE THIS
\usepackage{bbding}
\usepackage{booktabs}
\usepackage{multirow}
\usepackage{amsmath}
\usepackage{algorithm}
\usepackage{algorithmic}
\usepackage{url}
\usepackage[colorinlistoftodos,prependcaption,textsize=tiny]{todonotes}

\usepackage[colorinlistoftodos,prependcaption,textsize=tiny]{todonotes}

\usepackage{newfloat}
\usepackage{listings}
\floatstyle{ruled}
\newfloat{listing}{tb}{lst}{}
\floatname{listing}{Listing}
\iftrue
\title{Efficient AlphaFold2 Training using Parallel Evoformer and Branch Parallelism}
\author{
    Guoxia Wang\equalcontrib, Zhihua Wu\equalcontrib, Xiaomin Fang \\
    Yingfei Xiang, Yiqun Liu, Dianhai Yu, Yanjun Ma
}
\affiliations{
    Baidu Inc. \\
    mingzilaochongtu@gmail.com \\
    \{Wuzhihua02,fangxiaomin01,xiangyingfei01,liuyiqun01,yudianhai,mayanjun02\}@baidu.com
}
\fi

\usepackage{bibentry}
\begin{document}

\maketitle

\begin{abstract}
The accuracy of AlphaFold2, a frontier end-to-end structure prediction system, is already close to that of the experimental determination techniques. Due to the complex model architecture and large memory consumption, it requires lots of computational resources and time to train AlphaFold2 from scratch. Efficient AlphaFold2 training could accelerate the development of life science. In this paper, we propose a Parallel Evoformer and Branch Parallelism to speed up the training of AlphaFold2. We conduct sufficient experiments on UniFold implemented in PyTorch and HelixFold implemented in PaddlePaddle, and Branch Parallelism can improve the training performance by 38.67\% and 36.93\%, respectively. We also demonstrate that the accuracy of Parallel Evoformer could be on par with AlphaFold2 on the CASP14 and CAMEO datasets. The source code is available on \url{https://github.com/PaddlePaddle/PaddleFleetX}.
\end{abstract}

\begin{figure}[t]
\centering
\includegraphics[width=0.98\columnwidth]{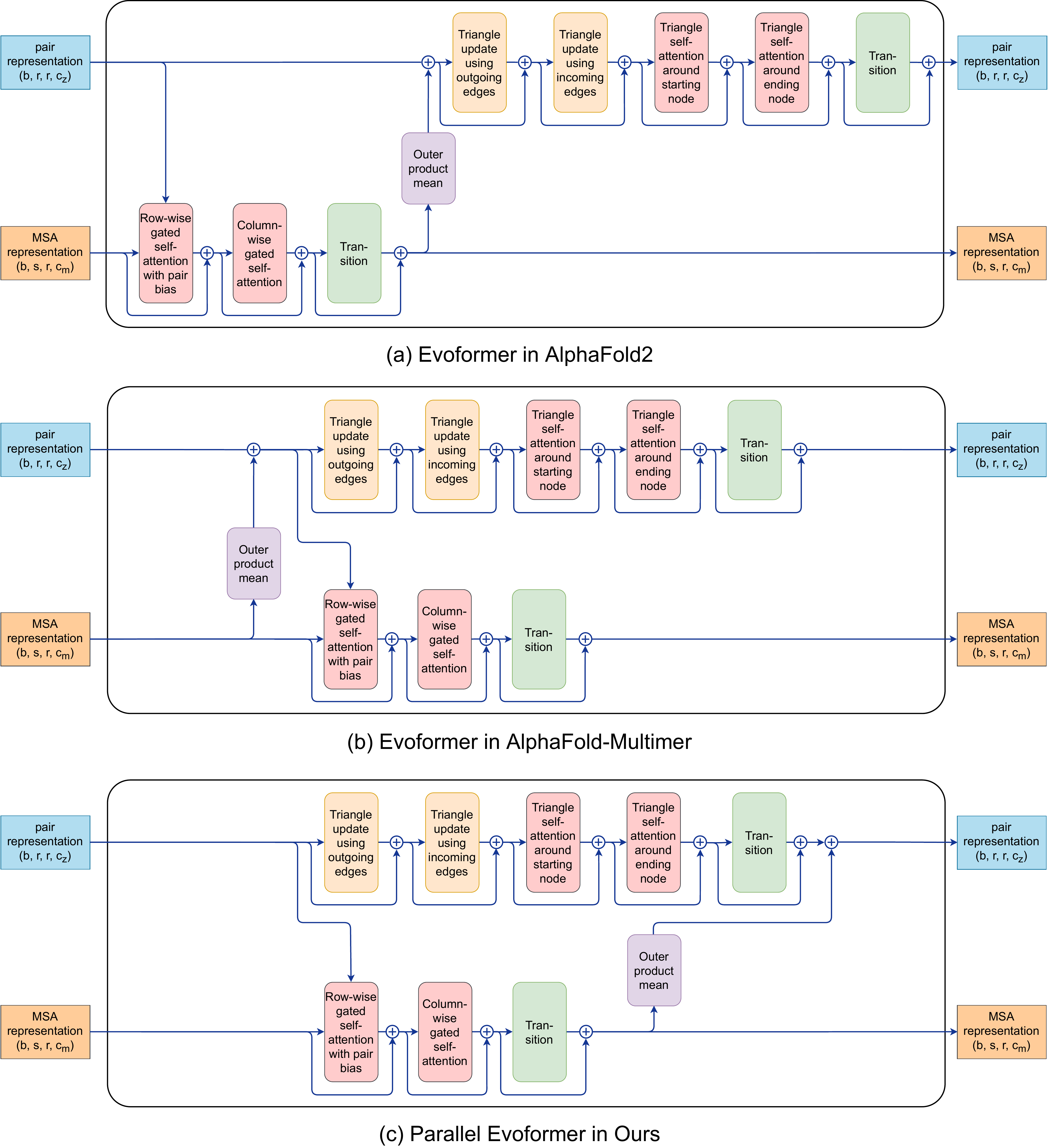} % Reduce the figure size so that it is slightly narrower than the column.
\caption{Various Evoformer block. (a) The original Evoformer block in AlphaFold2. (b) Modified Evoformer block in AlphaFold-Multimer. (c) The Parallel Evoformer block proposed in this paper. The main difference is that the outer product mean cross-communication happens at different position within the block.}
\label{fig:various_evoformer}
\end{figure}

\section{Introduction}
\begin{figure*}[t]
\centering
\includegraphics[width=0.98\textwidth]{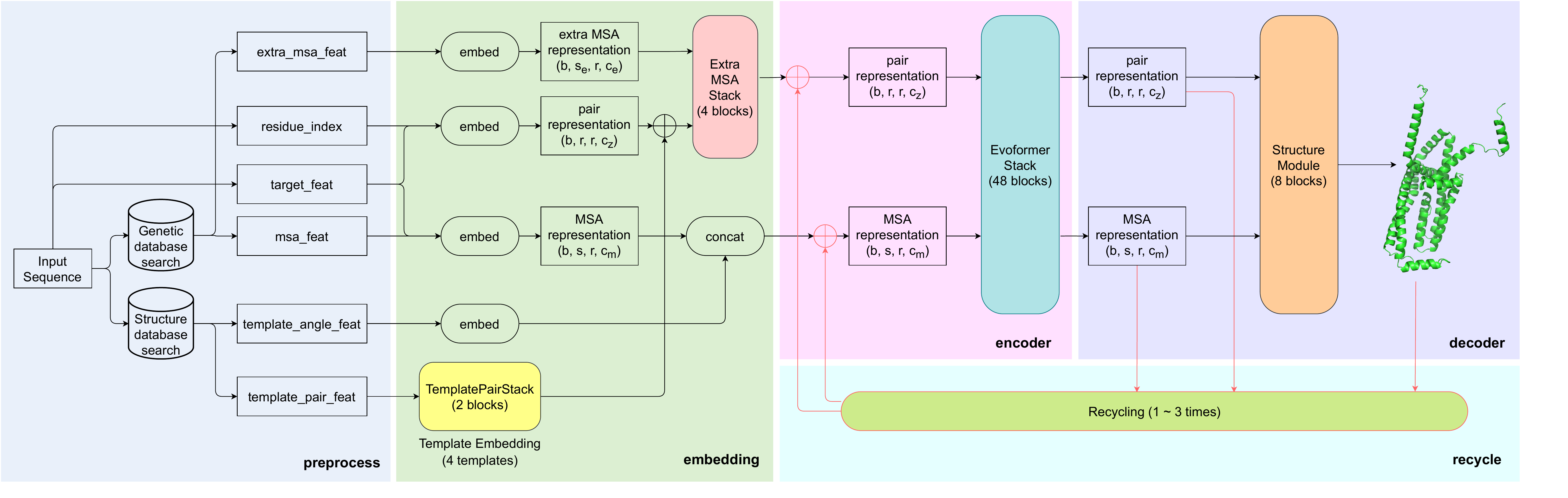} 
\caption{Overall framework of AlphaFold2. Dimension names: $b$: mini-batchs, $s$: clustered MSA sequences, $s_e$: extra MSA sequences, $r$: residues, $c$: channels. The Extra MSA stack is composed of Evoformer, so AlphaFold2 has 52 Evoformer blocks.}
\label{fig:alphafold2}
\end{figure*}

Proteins are exceptionally critical for life science, as it plays a wide range of functions in organisms. A protein comprises a chain of amino acid residues and folds into a 3D structure to play its functions. Since the 3D structure determines the protein's functions, studying the 3D structure helps to understand the mechanism of the protein's activities. However, it is time-consuming and complex to study protein structure determination through experimental technologies, e.g., X-ray crystallography and nuclear magnetic resonance (NMR). Until now, the experimental methods have determined about two hundred thousand protein structures \cite{sussman1998protein,10.1093/nar/gkaa1038}, only a fairly small portion of hundreds of millions of publicly available amino acid sequences \cite{10.1093/nar/gkw1099}. Therefore, efficient protein structure estimation methods are in great demand.

Many institutions \cite{jumper2021highly,yang2015tasser,du2021trrosetta,baek2021accurate,peng2011raptorx} made their efforts to develop AI-based protein structure prediction systems due to the efficiency and the capacity of the deep neural networks. In particular, thanks to the fantastic performance in the challenging 14th Critical Assessment of Protein Structure Prediction (CASP14) \cite{kryshtafovych2021critical}, AlphaFold2 \cite{jumper2021highly} from DeepMind has attracted lots of public attention. The accuracy of AlphaFold2 approaches that of the experimental determination technologies. AlphaFold2 is an end-to-end protein estimation pipeline that directly estimates the 3D coordinates of all the atoms in the proteins. A novel and well-designed architecture is proposed to promote the estimation accuracy, which jointly models multiple sequence alignments (MSAs) for evolutionary relationships and pairwise relations between the amino acids to learn the spatial relations. 

Although the accuracy of AlphaFold2 is satisfactory for protein structure prediction, it also takes 11 days to train end-to-end on 128 TPUv3 cores from scratch, limiting its wide usage. The structure of the AlphaFold2 is complex, as shown in Figure \ref{fig:alphafold2}, which leads to high training overhead. Specifically, there are three main reasons: First, the AlphaFold2 is relatively deep, and the Evoformer block has two computing branches and cannot be calculated in parallel. Second, the official open-source implemented total batch size is limited to 128, and each device has only 1 batch size, which cannot be extended to more devices in parallel to accelerate training through data parallelism. Third, although the parameters of AlphaFold2 are only 93M, the number of parameter tensors reaches 4630. The time overhead of accessing these small tensors in different training stages of each iteration is not negligible.

To this end, this paper proposes two optimization techniques for two of the above three problems to achieve efficient AlphaFold2 training under the premise of fully aligning hyperparameters (network model configuration and total batchsize of 128 with 1 protein sample per device). First, inspired by AlphaFold-mutimer \cite{Evans2021Protein}, we modify the two serial computing branches in the Evoformer block into a parallel computing structure, named Parallel Evoformer, as shown in Figure \ref{fig:various_evoformer}. Second, we propose a novel Branch Parallelism (BP) for Parallel Evoformer, which can break the barrier of parallel acceleration that cannot be scaled to more devices through data parallelism due to a batch size of 1 on each device.

The method proposed in this paper to efficiently train AlphaFold2 models is general and not limited to deep learning frameworks and the version of re-implemented AlphaFold2. We perform extensive experimental verification on UniFold implemented in PyTorch and HelixFold implemented in PaddlePaddle. Extensive experimental results show that Branch Parallelism can achieve similar training performance improvements on both UniFold and HelixFold, which are 38.67\% and 36.93\% higher, respectively. We also demonstrate that the accuracy of Parallel Evoformer could be on par with AlphaFold2 on the CASP14 and CAMEO datasets.

The main contributions of this paper can be summarized as follows:
\begin{itemize}
    \item We improve the Evoformer in AlphaFold2 to Parallel Evoformer, which breaks the computational dependency of MSA and pair representation, and experiments show that this does not affect the accuracy.
    \item We propose Branch Parallelism for Parallel Evoformer, which splits different computing branches across more devices in parallel to speed up training efficiency. This breaks the limitation of data parallelism in the official implementation of AlphaFold2.
    \item We reduce the end-to-end training time of AlphaFold2 to 4.18 days on UniFold and 4.88 days on HelixFold, improving the training performance by 38.67\% and 36.93\%, respectively. It achieves efficient AlphaFold2 training, saving R\&D economic costs for biocomputing research.
\end{itemize}

\section{Background}

\subsection{Overview of AlphaFold2}
Comparing to the traditional protein structure prediction model which usually consists of multiple steps, AlphaFold2 processes the input protein sequence and predicts the 3D protein structure through an end-to-end procedure. In general, AlphaFold2 takes the amino acid sequence as input and then search against protein databases to obtain MSAs and similar templates. By using MSA information, we can detect correlations between the parts of similar sequences that are more likely to mutate. The templates with regards to the input sequence, on the other hand, provide structural information for the model to predict the final structure.

The overall framework of AlphaFold2 can be divided into five parts: Preprocess, Embedding, Encoder, Decoder, and Recycle, which is shown in Figure \ref{fig:alphafold2}. The Preprocess part mainly parses the input raw sequence and generates MSA-related and template-related features via genetic database search and structure database search. The features are then embedded into MSA representation, pair representation and extra MSA representation during Embedding part. These representations contain sufficient co-evolutionary information among similar sequences and geometric information of residue pairs within the sequence. The third part consists of 48-layer Evoformer blocks that iteratively refine and exchange information between the MSA representation and the pair representation. After obtaining the refined representations, the 8-block Structure Module acts as a decoder to directly generate the final structure. Moreover, AlphaFold2 adopts Recycling technique to improve the accuracy of prediction by passing the representations through Evoformer and Structure Module repeatedly.

\subsection{Evoformer}
Evoformer is the core module of AlphaFold2, which includes two tracks that handle MSA representation and pair representation, as well as the communication scheme between them. As shown in Figure \ref{fig:various_evoformer}(a), MSA representation is processed with Row-wise gated self-attention with pair bias, Column-wise gated self-attention and Transition, while pair representation is further processed with Triangle update, Triangle self-attention and Transition. The outer product mean is used to pass the information between the two tracks.

\section{Related Work}
% \begin{figure}[t]
% \centering
% \includegraphics[width=0.96\columnwidth]{parallelisms} % Reduce the figure size so that it is slightly narrower than the column.
% \caption{Various Parallelism Techniques. (a) Tensor Model Parallelism splits the parameter into individual layers to multiple devices at the tensor dimension instead of the input activation; (b) Pipeline Model Parallelism splits the model at the layer dimension; (c) Dynamic Axial Parallelism splits the input activation rather than the model parameter, e.g. the split MSA representation shape is $[B, N_{seq}/N_{device}, N_{res}, C_m]$ in Evoformer; (d) Branch Parallelism splits the calculation branch to multiple devices where each device has the same input activation and parameter.}
% \label{fig_parallelisms}
% \end{figure}

\subsection{Evoformer Improvment for AlphaFold2}
Although many biological or pharmaceutical studies are based on AlphaFold2 after its open source in July 2021, few works focused on improving the main modules of AlphaFold2. ParaFold \cite{10.1145/3503470.3503471} aims to accelerate the inference pipeline of AlphaFold2 by performing MSA searches with multiprocessing and distributing different procedures within Evoformer and Structure Module to CPUs or GPUs, instead of optimizing the network structure of AlphaFold2. In order to predict larger proteins, AlphaFold-Multimer \cite{Evans2021Protein} makes minor architectural modifications of AlphaFold2, including swapping the attention and triangular multiplicative update layers in the template stack which aligns the order in the Evoformer stack, and moving the outer product mean to the start of the Evoformer block which ensures MSA stack and pair stack can be processed in parallel. Recently, several works proposed new training schemes and improved main structures of AlphaFold2 which enable the prediction of proteins accurately from single primary sequence alone. Meta's ESMFold \cite{Lin2022.07.20.500902} replaces the axial attention with a standard attention in order to adapt to Evoformer block. OmegaFold from HeliXon \cite{Wu2022.07.21.500999} optimizes Evoformer with simplified node attention and edge attention to capture complex interaction patterns among amino acids. HelixFold-Single \cite{fang2022helixfold_single} designs an adaptor to align the output of pretrained protein language model to Evoformer and the column-wise gated self-attention is removed due to no necessity of exchanging the messages within the MSAs.

\subsection{Distributed Parallel Strategy}
\begin{figure}[t]
\centering
\includegraphics[width=0.99\columnwidth]{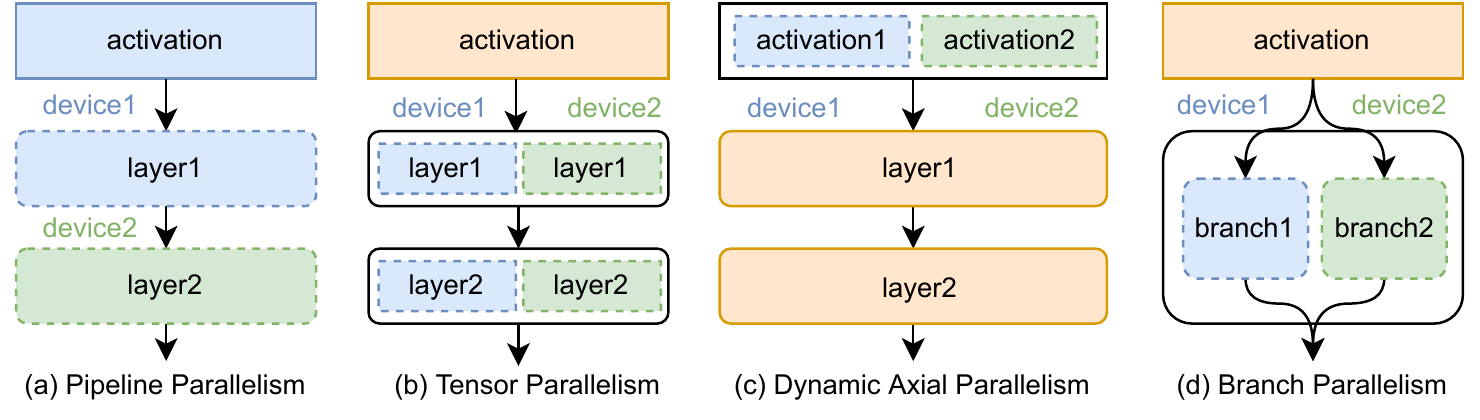} % Reduce the figure size so that it is slightly narrower than the column.
\caption{Various distributed parallel strategies.}
% \caption{Various Parallelism Techniques. (a) Tensor Model Parallelism splits the parameter into individual layers to multiple devices at the tensor dimension instead of the input activation; (b) Pipeline Model Parallelism splits the model at the layer dimension; (c) Dynamic Axial Parallelism splits the input activation rather than the model parameter, e.g. the split MSA representation shape is $[B, N_{seq}/N_{device}, N_{res}, C_m]$ in Evoformer; (d) Branch Parallelism splits the calculation branch to multiple devices where each device has the same input activation and parameter.}
\label{fig_parallelisms}
\end{figure}

To improve training speed, data parallelism (DP) \cite{li2020pytorch} is the most popular and efficient method in deep learning distributed training. Each worker has a copy of the model parameters, and parallelism operates in the data dimension. Each device accepts mini-batch samples while scaling to more devices at the expense of increasing the total batch size. DeepSpeed’s ZeRO \cite{rajbhandari2020zero} and FairScale’s Fully Sharded Data Parallel \cite{FairScale2021} reduce redundant storage of tensors through communication costs. Model parallelism (MP) \cite{narayanan2021efficient} uses more accelerators to train large-scale models, which can be divided into pipeline model parallelism (PP) and tensor model parallelism (TP). PP splits the whole model to multiple devices at layer dimension and will introduce the idle time named \textit{pipeline bubble}, and TP distributes the parameter in individual layers to multiple devices at the tensor dimension. Dynamic axial parallelism (DAP) \cite{cheng2022fastfold} is proposed to solve the inefficient problem of training the AlphaFold2 model with a small parameter shape and a large activation memory consumption. Our proposed branch parallelism is different from other parallel computing methods. For the feature of two computing branches in Evoformer block, BP splits the two computing branches of MSA stack and pair stack into different devices for parallel computing. The difference between the parallel methods is shown in Figure \ref{fig_parallelisms}.

% \subsection{Fusion Technology}
% The goal of fusion technology is to reduce the additional overhead and increase computational density. Operator fusion combines multiple small operators into a large operator, reducing the CPU scheduling overhead between small operators. Similarly, tensor fusion fuses multiple small tensors into one large tensor, which improves memory access efficiency while reducing the CPU scheduling overhead of traversing multiple small tensors. Many Transformer-based systems \cite{rasley2020deepspeed,wang2021lightseq2,fang2021turbotransformers} attempt to fuse the performance-critical multi-head attention (MHA) module into a single operator to increase the throughput. We fuse the Gated Self-Attention consisting of 14 small operators in the Evoformer block into one operator. Distributed data parallelism fuse multiple gradients to a large tensor and synchronizes them by \emph{Allreduce} operation which is implemented in PyTorch \cite{NEURIPS2019_9015} and PaddlePaddle \cite{ma2019paddlepaddle}. Approaches like DeepSpeed’s ZeRO \cite{rajbhandari2020zero} and FairScale’s Fully Sharded Data Parallel \cite{FairScale2021} fuse not only gradients but also parameters and optimizer states. We use ultimate tensor fusion in AlphaFold2 to improve performance, fusing 4630 small tensors into 1 large tensor and using the fused large tensor for computation in different stages.

\section{Implementation}

\subsection{Parallel Evoformer}

Evoformer of AlphaFold2 has two computational branches with axial self-attention in the MSA stack; triangular multiplicative updates and triangular self-attention in the pair stack; and an information exchange mechanism that outer product mean and attention biasing to allow communication between the stacks, as shown in Figure \ref{fig:various_evoformer}(a). 
The two computing branches of the original Evoformer are serial computing, where the input of the pair stack depends on the output of the MSA stack. AlphaFold-Multimer \cite{Evans2021Protein} moves the outer product mean to the start of the Evoformer block. Its main idea is to save the activation memory of the MSA stack and pair stack during backpropagation at training time and to process both stacks in parallel at inference time, as shown in Figure \ref{fig:various_evoformer}(b). However, the calculation of the pair stack still depends on the output of the outer product mean, which leaves room for improvement in parallel efficiency. In order to achieve independent computing and fully parallel computing of the two branches, we propose to move the outer product mean to the end of the Evoformer block, named Parallel Evoformer, see in Figure \ref{fig:various_evoformer}(c). The main trunk of the AlphaFold2 consists of 52 (48+4) Evoformer blocks. 
Instead of moving the outer product mean to the start of the Evoformer block, Parallel Evoformer still allows the pair representation and the MSA representation to evolve independently within a given block, with all cross communication happening at the end of the block. Since the overall process across the Evoformer block is learnable, the position of the outer product mean doesn't affect the accuracy of the final prediction as the number of the Evoformer blocks increases, which is further proved by our experiments shown in Figure \ref{fig:evoformer_log}.

\subsection{Branch Parallelism}

\begin{figure}[t]
\centering
\includegraphics[width=0.98\columnwidth]{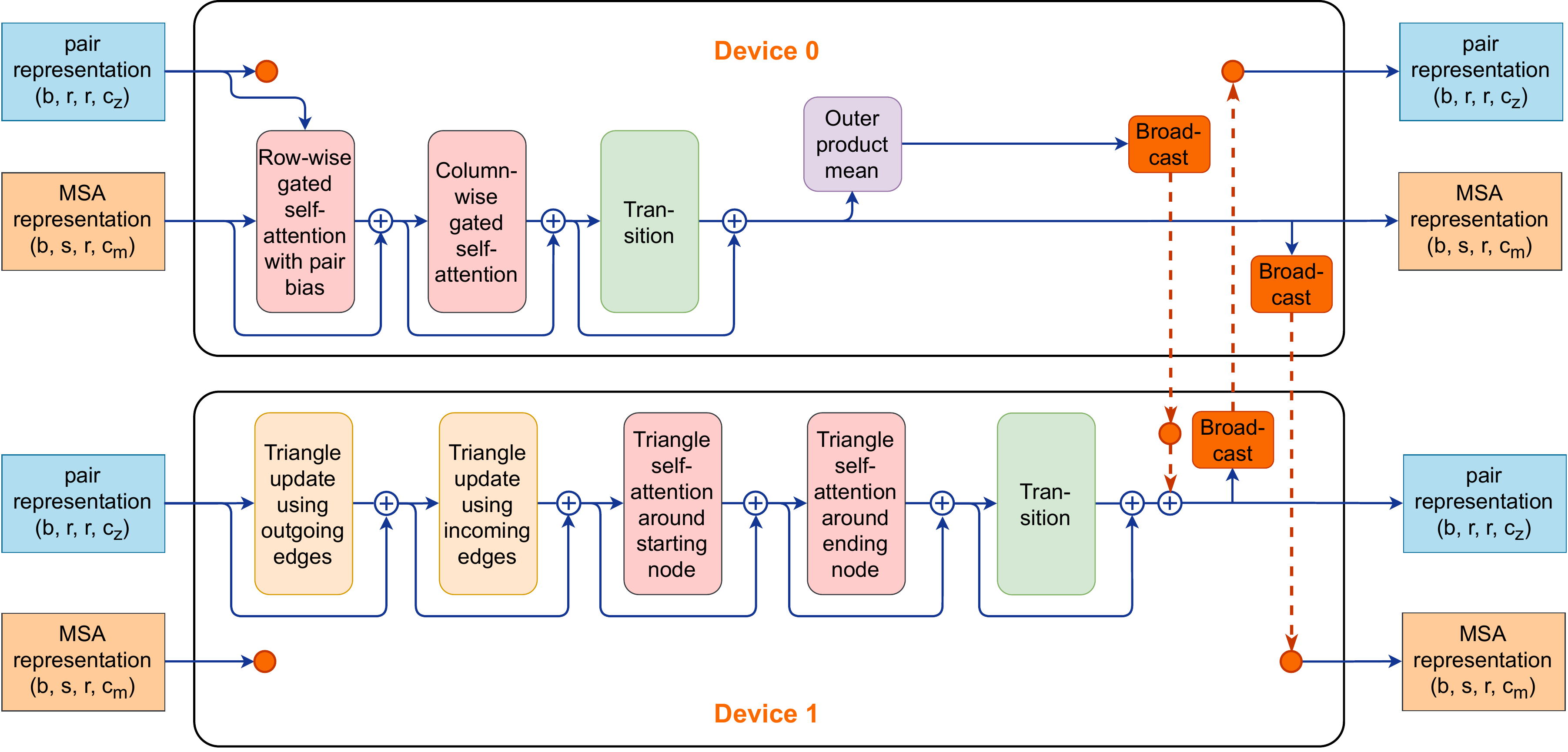} % Reduce the figure size so that it is slightly narrower than the column.
\caption{Branch parallelism implementation of Parallel Evoformer. Branch parallelism splits Parallel Evoformer's two computing branches across two devices. The two branches are calculated in parallel and the results are synchronized through the broadcast communication operator.}
\label{fig:evoformer_bp_eachrank}
\end{figure}

In order for the two computing branches in Parallel Evoformer to be computed in parallel, we propose Branch Parallelism (BP), and Figures \ref{fig:evoformer_bp_eachrank} shows the details. BP is a novel parallel technique, which can be applied not only to the AlphaFold2 Evoformer model structure, but also to a model structure with multiple parallel computing branches and an approximate amount of computation. BP splits the calculation branch across multiple devices. One device calculates the MSA stack and the other calculates the pair stack. To preserve strict calculation semantics, BP inserts \emph{Broadcast} to and \emph{AllReduce} communication. 

Specifically, in the forward stage, we do not need to split the input in each Evoformer block as the first device uses the MSA and pair representation to calculate the MSA stack and outer product mean branch then the outputs are broadcasted to the second device. The second device uses the pair representation to calculate the pair stack branch, adds the outer product mean broadcasted from the first device, and then broadcasts the output to the first device. In the backward stage, we insert \emph{Broadcast} to synchronize the gradient of the outer product mean from the second device and \emph{AllReduce} to accumulate the sum of the gradient of the input pair representation in each Evoformer block. At the end of backward propagation of the whole Evoformer module, we need an extra broadcast of the gradient of MSA representation. Finally, we adopt the \emph{AllReduce} or \emph{Broadcast} communication to synchronize the gradients of Evoformer model parameters.

BP does not split the intermediate activation for the Evoformer block dominated by small kernels, as a result, the same computational intensity is retained. At the same time, BP can compute the two branches of Evoformer completely in parallel. However, BP also has its limitations: it can be scaled up to as many devices as the number of computational branches and requires a similar amount of computation from each branch.

\subsection{Hybrid Parallelism}

DAP \cite{cheng2022fastfold} splits the activation to reduce the memory footprint. In addition, DAP also has high parallel efficiency with a large shape of input, e.g. in the fine-tuning stage. Therefore, we can combine BP and DAP to train the AlphaFold2 model when CPU launch overheads are not the performance bottleneck. Since BP and DAP just split the computing branches and the activation across multiple devices respectively, the same protein is processed on each device. To improve communication efficiency, BP and DAP work within a single node where the inter-GPU communication bandwidth is high. With data parallelism, we can scale the total mini-batch to 128. We call this technique BP-DAP-DP hybrid parallelism.

\section{Experiments}

\subsection{Experimental Setup}

\subsubsection{Third-party implementation of AlphaFold2}
Since the original AlphaFold2 only open sourced the inference code but not the training code, multiple teams have re-implemented and optimized AlphaFold2 training on different deep learning frameworks. We perform experimental verification on UniFold \cite{uni-fold} implemented in PyTorch \cite{NEURIPS2019_9015} and HelixFold \cite{wang2022helixfold} implemented in PaddlePaddle \cite{ma2019paddlepaddle}.

\subsubsection{Datasets}
For training, we follow the settings reported in the paper of AlphaFold2 to collect the training data, including 25\% of samples from RCSB PDB (\url{https://www.rcsb.org/}) \cite{10.1093/nar/28.1.235,10.1093/nar/gkaa1038} and 75\% of self-distillation samples. For evaluation, we collect two datasets: CASP14 and CAMEO. We collect 87 domain targets from CASP14 (\url{https://predictioncenter.org/casp14/index.cgi}) \cite{jumper2021highly,https://doi.org/10.1002/prot.26202,https://doi.org/10.1002/prot.26237}. We also collect 371 protein targets from CAMEO (\url{https://www.cameo3d.org/}) \cite{https://doi.org/10.1002/prot.26213}, ranging from 2021-09-04 to 2022-02-19.

\subsubsection{Settings of Model Architectures}
We use two model settings to assess the AlphaFold2 model training speed improved by this paper. The model settings are shown in Table \ref{tab:model_settins}, with \emph{initial training} setting corresponding to model 1 and \emph{fine-tuning} setting corresponding to model 1.1.1 reported in the supplementary information of paper AlphaFold2.

\subsubsection{Hyperparameter setting}
All our experiments are run on NVIDIA A100 (40G) and the mini-batch size is 1 on each device. We strictly follow the settings in AlphaFold2. As a feature in AlphaFold2, recycling iteration is to randomize a number from 1 to 4 in each step, performs the forward pass multiple times, and then performs the backward pass. To compare performance quickly and fairly, we firstly fix the random seed, then run 105 training steps, discard the first 5 steps, and finally calculate the average 
speed for the last 100 steps. After our extensive experimental verification, the average speed of 100 steps can get a similar global speed. We fix $random\_seed = 32$, then the random seed for each step is calculated by $random\_seed + cur\_step$. Unless otherwise specified, we use AMP for training, using the Float32 parameter and BFloat16 intermediate activation.

\begin{table}[t]
\centering
\resizebox{.95\columnwidth}{!}{
\begin{tabular}{cccccc}
   \toprule
   Training Process & Model setting & $N_{templ}$ & $N_{res}$ & $N_{seq}$ & $N_{extra\_seq}$\\
   \midrule
   Initial training & Model 1 & 4 & 256 & 128 & 1024 \\
   Fine-tuning & Model 1.1.1 & 4 & 384 & 512 & 5120 \\
   \bottomrule
\end{tabular}
}
\caption{Model settings for performance comparison.}
\label{tab:model_settins}
\end{table}

\subsection{Effectiveness of Parallel Evoformer}
\begin{figure}[t]
\centering
\includegraphics[width=0.98\columnwidth]{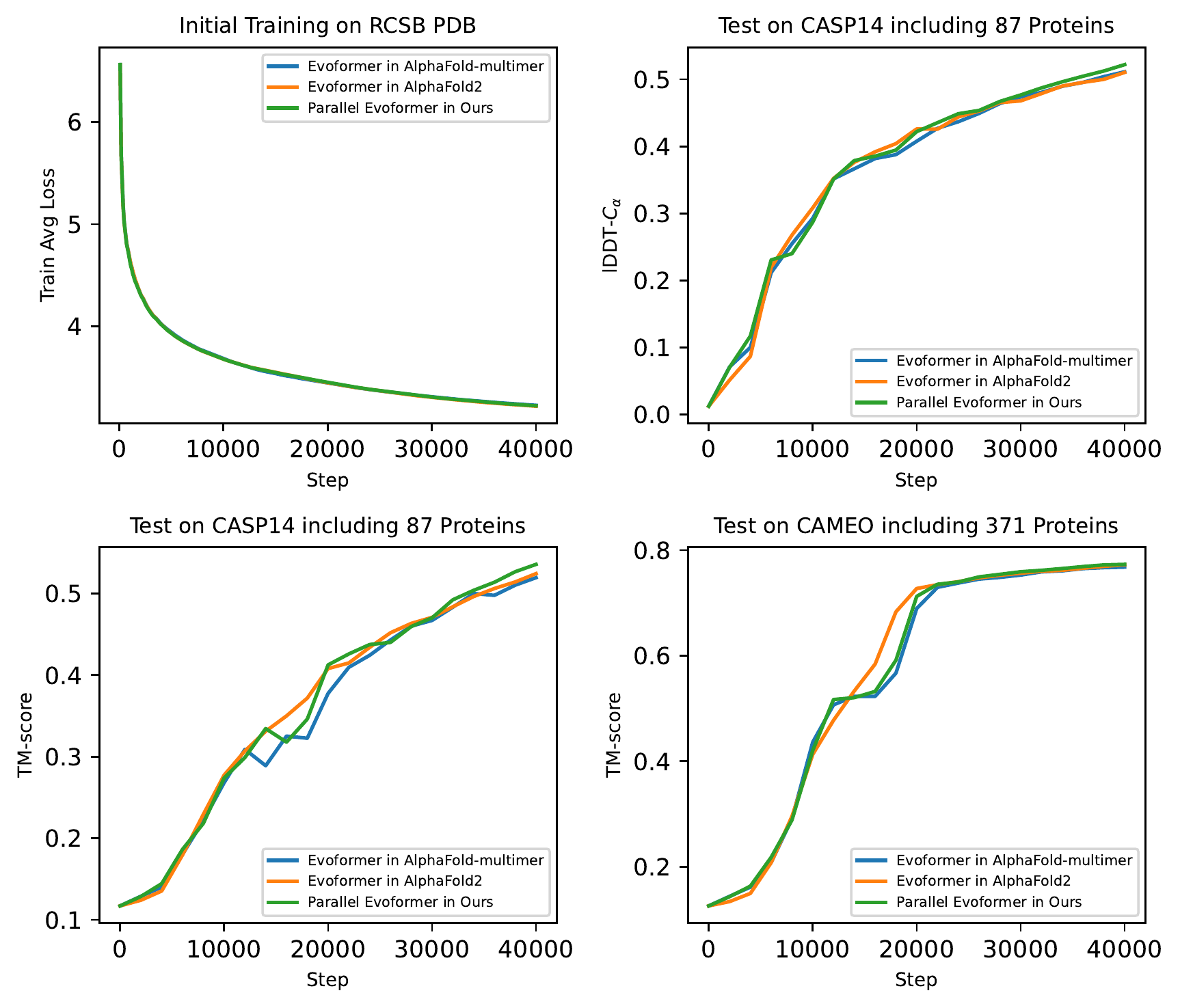} % Reduce the figure size so that it is slightly narrower than the column.
\caption{Accuracy comparison of various Evoformer block on HelixFold.}
\label{fig:evoformer_log}
\end{figure}

To demonstrate the effectiveness of the Evoformer block modification, we test 3 different Evoformer blocks shown in Figure \ref{fig:various_evoformer}. We train it from scratch on a single node with 8 A100 GPUs and the batch size is 1 per GPU. The training dataset consists of 120,000 proteins from the RCSB PDB, the learning rate is 5e-5, the global gradient clipping value is 0.1, the number of warm-up steps is 1000 and total training step is 40,000. Figure \ref{fig:evoformer_log} shows the training loss, TM-score and lDDT-C$_\alpha$ metrics on the CASP14 and CAMEO test sets. The results show that Parallel Evoformer can achieve competitive accuracy with Evoformer in AlphaFold2 and AlphaFold-Mutimer. The training speed of the 3 different Evoformer blocks is the same. We also report the computational overhead ratio for 52 (48+4) Evoformer blocks, as shown in Table \ref{tab:evoformer_speed}. It means that the position of the outer product mean does not affect the accuracy and training speed of AlphaFold2.

\subsection{Training Performance}

\begin{table}[t]
\centering
\resizebox{.98\columnwidth}{!}{
\begin{tabular}{*{6}{c}}
   \toprule
   \multirow{2}{*}{Training Process} & \multirow{2}{*}{Evoformer} & \multicolumn{4}{c}{s/step} \\
   \cmidrule(lr){3-6}
    &  & Evoformer & Other & Total & (\%) \\
   \midrule
   \multirow{3}{*}{Initial training} & AlphaFold2 & 3.12 & 1.88 & 5.00 & 62.40\% \\
  ~ & AlphaFold-Multimer & 3.09 & 1.89 & 4.98 & 62.04\% \\
  ~ & \textbf{Parallel Evoformer} & 3.11 & 1.89 & 5.00 & 62.20\% \\
   \midrule
   \multirow{3}{*}{Fine-tuning} & AlphaFold2 & 12.81 & 3.69 & 16.50 & 77.63\% \\
  ~ & AlphaFold-Multimer & 12.86 & 3.65 & 16.51 & 77.89\% \\
  ~ & \textbf{Parallel Evoformer} & 12.83 & 3.67 & 16.50 & 77.75\% \\
   \bottomrule
\end{tabular}
}
\caption{The effect of modification of the Evoformer block on training performance on HelixFold. The ratio of Evoformer block computation time to total time cost can help to understand the performance improvement of Branch Parallelism. The experiments are performed on 8 A100 GPUs, and CUDA synchronization operations are added.}
\label{tab:evoformer_speed}
\end{table}

\begin{table}[t]
\centering
\resizebox{.95\columnwidth}{!}{
\begin{tabular}{ccccccc}
   \toprule
   Implementation & Training Process & Dtype & BP & s/step & protein/s & (\%)  \\
   \midrule
   \multirow{2}{*}{UniFold} & \multirow{2}{*}{Initial training} & FP32 & 1 & 7.04 & 18.18 & - \\
   ~ & ~ & FP32 & 2 & 5.41 & 23.65 & +30.12\%  \\
   \midrule
   \multirow{2}{*}{HelixFold} & \multirow{2}{*}{Initial training} & FP32 & 1 & 6.48 & 19.75 & -  \\
   ~ & ~ & FP32 & 2 & 4.55 & 28.13 & +42.41\%  \\
   \midrule
   \multirow{4}{*}{UniFold} & \multirow{2}{*}{Initial training} & BF16 & 1 & 4.16 & 30.76 & -  \\
   ~ & ~ & BF16 & 2 & 3.02 & 42.38 & \textbf{+37.74\%}  \\
   \cmidrule(lr){3-7}
   ~ & \multirow{2}{*}{Fine-tuning} & BF16 & 1 & 15.02 & 8.52 & -  \\
   ~ & ~ & BF16 & 2 & 10.70 & 11.96 & \textbf{+40.37\%} \\
   \midrule
   \multirow{4}{*}{HelixFold} & \multirow{2}{*}{Initial training} & BF16 & 1 & 4.92 & 26.01 & -  \\
   ~ & ~ & BF16 & 2 & 3.55 & 36.05 & \textbf{+38.59\%}  \\
   \cmidrule(lr){3-7}
   ~ & \multirow{2}{*}{Fine-tuning} & BF16 & 1 & 16.45 & 7.78 & -  \\
   ~ & ~ & BF16 & 2 & 12.29 & 10.41 & \textbf{+33.84\%}  \\
   \bottomrule
\end{tabular}
}
\caption{Performance improvements for Branch Parallelism on HelixFold and UniFold. The total batch size of the test is 128, with 1 protein sample per GPU. Fine-tuning performance is not reported because fine-tuning of Float32 data type can cause OOM on A100 40G.}
\label{tab:bp_performance}
\end{table}

\subsubsection{Performance Comparison of BP and DAP}

The author of FastFold \cite{cheng2022fastfold} has open-sourced DAP. To compare with the performance of DAP, we firstly use PaddlePaddle to reproduce the open-source code of FastFold, called PPFold. Then we add the implementation of BP on PPFold to compare the performance of DAP and BP.

Table \ref{tab_pp_speed} shows the performance comparison of FastFold and PPFold. As can be seen in the table, DAP=2 uses 2 GPUs compared to DAP=1, but there is a performance drop in both FastFold and PPFold as the last two dimensions of the inputs are relatively small with a low computational intensity. DAP splits the input into smaller tensors, and the computational intensity is not improved. In addition, a lot of additional communication overhead is introduced, resulting in a decrease in performance. BP 
splits the computing branches across different GPUs, which can be calculated in parallel while maintaining computational intensity, and only introducing a small amount of communication overhead. Thus, the performance is improved by 67.45\%. Similar performance is also observed in end-to-end training, see Table \ref{tab_hybrid_speed}.

These require an extra declaration. This paper does not compare BP with other distributed parallelisms such as TP and PP. There are two reasons for this. First, FastFold has compared DAP, TP and PP on AlphaFold2. The experimental results show that the acceleration efficiency of DAP is higher than that of TP and PP. Second, implementing TP and PP on AlphaFold2 requires a lot of work.
\begin{table*}[t]
\centering
\resizebox{0.99\textwidth}{!}{
\begin{tabular}{ccccccccc}
   \toprule
   Implementation & Training Process & Hardware & Step Time (s) & Protein/s & Training Time (days) & Total (days) & (\%) \\
   \midrule
   \multirow{2}{*}{AlphaFold2-DP} & Initial training & \multirow{2}{*}{128 × TPUv3} & 7.513 & 17.037 & 6.793 & \multirow{2}{*}{10.961} \\
   ~ & Fine-tuning & ~ & 30.729 & 4.165 & 4.167 & ~ \\
   \midrule
   \multirow{2}{*}{OpenFold-DP} & Initial training & \multirow{2}{*}{128 × A100(40G)} & 8.9 & 14.382 & 8.047 & \multirow{2}{*}{10.849} \\
   ~ & Fine-tuning & ~ & 20.657 & 6.196 & 2.801 & ~ \\
   \midrule
   \multirow{2}{*}{UniFold-DP} & Initial training & \multirow{2}{*}{128 × A100(40G)} & 4.16 & 30.76 & 3.761 & \multirow{2}{*}{5.798} & \multirow{2}{*}{-} \\
   ~ & Fine-tuning & ~ & 15.02 & 8.52 & 2.037 & ~ \\
   \midrule
   \multirow{2}{*}{UniFold-BP} & Initial training & 256 × A100(40G) & 3.02 & 42.38 & 2.730 & \multirow{2}{*}{ \textbf{4.181}} & \multirow{2}{*}{\textbf{+38.67\%}} \\
   ~ & Fine-tuning & 256 × A100(40G) & 10.70 & 11.96 & 1.451 & ~ \\
   \midrule
   \multirow{2}{*}{HelixFold-DP$^\dagger$} & Initial training & \multirow{2}{*}{128 × A100(40G)} & 4.925 & 25.989 & 4.453 & \multirow{2}{*}{6.685} & \multirow{2}{*}{-} \\
   ~ & Fine-tuning & ~ & 16.458 & 7.777 & 2.232 & ~ \\
   \midrule
   \multirow{2}{*}{HelixFold-BP} & Initial training & 256 × A100(40G) & 3.555 & 36.005 & 3.214 & \multirow{2}{*}{ \textbf{4.882}} & \multirow{2}{*}{\textbf{+36.93\%}} \\
   ~ & Fine-tuning & 256 × A100(40G) & 12.298 & 10.407 & 1.668 & ~ \\
   \bottomrule
\end{tabular}
}
\caption{Complete end-to-end training performance. In the initial training stage, we train $10 \times 10^6$ samples (78125 steps), and in the fine-tuning stage, we continue to train $1.5 \times 10^6$ samples (11718 steps). *-DP and *-BP refer to using only data parallelism and a hybrid of data parallelism and branch parallelism, respectively. UniFold results are based on its GitHub commit 726480e.}
\label{tab_end2end_performance}
\end{table*}

\subsubsection{Performance of Branch Parallelism}

Branch Parallelism is a general distributed parallelism strategy that can be applied to AlphaFold2 models implemented by different deep learning frameworks, such as UniFold implemented in PyTorch and HelixFold implemented in PaddlePaddle. We conduct extensive experiments with Float32 and BFloat16 data types, in the mode of initial training and fine-tuning, on UniFold and HelixFold, respectively. As shown in Table \ref{tab:bp_performance}, the performance improvement of BP on Float32 is lower than that on BFloat16, but both still have more than 30\% performance speedups. In general, on UniFold and HelixFold, two different AlphaFold2 implementations both have similar performance speedups. With the default BFloat16 data type, UniFold improves by 37.74\% and 40.37\% in the initial training and fine-tuning training stages, respectively. Similarly, HelixFold also achieved 38.59\% and 33.84\% performance speedup respectively. The performance improvement is lower than about 40\%. The main reason is that the main module of the AlphaFold2 model is the Evoformer block, but there are other module computing overheads, as shown in Table \ref{tab:evoformer_speed}. Branch Parallelism is only calculated in parallel in the two branches of the Evoformer block.

\begin{table}[t]
\centering
\resizebox{.98\columnwidth}{!}{
\begin{tabular}{ccccc}
   \toprule
   Method & DAP & BP & Fwd + Bwd Time / Layer (ms) &  \\
   \midrule
   \multirow{2}{*}{FastFold} & 1 & 1 & 30.98 & - \\
   ~ & 2 & 1 & 32.25 & -3.94\% \\
   \midrule
   \multirow{3}{*}{PPFold} & 1 & 1 & 32.47 & - \\
   ~ & 2 & 1 & 33.21 & -2.22\% \\
   ~ & 1 & 2 & 19.39 & +67.45\% \\
   \bottomrule
\end{tabular}
}
\caption{FastFold VS PPFold performance comparison. Compare the total time of forward computation and backward computation for each layer. 12-layer Evoformer, data type is Float16, $head=8,B=1,N_{seq}=128,N_{res}=256,C_m=256,C_z=128$, the settings are the same as FastFold open source code. PPFold does not use asynchronous communication in DAP.}
\label{tab_pp_speed}
\end{table}

\subsubsection{Performance of Hybrid Parallelism}
% The author of FastFold \cite{cheng2022fastfold} has open-sourced DAP, but the open-source code only has the parallel implementation of the Evoformer module, and the related code implementations such as \emph{TemplatePairStack} and \emph{StructureModule} are not available to the public. To compare with the performance of DAP, we firstly use PaddlePaddle to reproduce the open-source code of FastFold, called PPFold. Then we add the implementation of BP on PPFold to compare the performance of DAP and BP.

% Table \ref{tab_pp_speed} shows the performance comparison of FastFold and PPFold. As can be seen in the table, DAP=2 uses 2 GPUs compared to DAP=1, but there is a performance drop in both FastFold and PPFold as the last two dimensions of the inputs are relatively small with a low computational intensity. DAP splits the input into smaller tensors, and the computational intensity is not improved. In addition, a lot of additional communication overhead is introduced, resulting in a decrease in performance. BP 
% splits the computing branches across different GPUs, which can be calculated in parallel while maintaining computational intensity, and only introducing a small amount of communication overhead. Thus, the performance is improved by 67.45\%.

To illustrate the effectiveness of BP combined with other parallel strategies, we conduct experiments with different configurations using hybrid parallelism on HelixFold. As shown in Table \ref{tab_hybrid_speed}, in the initial training, where the dimensions involved in the computation are relatively small, the performance of DAP=2 drops compared to that of unused, showing a negative gain. When BP=2, the performance is improved by 38.51\%, indicating a positive benefit. Conversely, in the fine-tuning, where the dimensions of MSA depth and the length of amino acid sequences increase, DAP achieves a higher performance improvement than BP by splitting larger activations across multiple GPUs for parallel computing. However, the hybrid parallelism of DAP=2 and BP=2 has the approximate throughput of DAP=4 and BP=1 while DAP=4 and BP=2 have higher throughput than DAP=8 and BP=1, demonstrating that when the activation is divided to a certain size, the parallelism efficiency of BP is higher than that of DAP.

\begin{table}[t]
\centering
\resizebox{.98\columnwidth}{!}{
\begin{tabular}{cccccc}
   \toprule
   Training Process & DAP & BP & s/step & protein/s & (\%) \\
   \midrule
   \multirow{3}{*}{Initial training} & 1 & 1 & 4.925 & 25.989 & - \\
   ~ & 2 & 1 & 5.170 & 24.758 & -4.73\% \\
   ~ & 1 & 2 & 3.555 & 36.005 & +38.51\% \\
   \midrule
   \multirow{6}{*}{Fine-tuning} & 1 & 1 & 16.458 & 7.777 & - \\
   ~ & 2 & 1 & 11.110 & 11.521 & +48.13\% \\
   ~ & 1 & 2 & 12.298 & 10.408 & +33.82\% \\
   ~ & 2 & 2 & 7.887 & 16.229  & \textbf{+108.66\%} \\
   ~ & 4 & 1 & 7.883 & 16.237  & \textbf{+108.77\%} \\
   ~ & 8 & 1 & 7.315 & 17.498  & +124.97\% \\
   ~ & 4 & 2 & 5.700 & 22.456  & \textbf{+188.71\%} \\
   \bottomrule
\end{tabular}
}
\caption{Performance on HelixFold with hybrid parallelism. The total batch size is 128, with 1 protein sample per GPU.}
\label{tab_hybrid_speed}
\end{table}

\subsubsection{End-to-end training Training Performance}

We make a comparison among AlphaFold2 \cite{jumper2021highly}, OpenFold \cite{Ahdritz_OpenFold_2021}, HelixFold \cite{wang2022helixfold} and UniFold \cite{uni-fold}, in terms of hardware, time cost of each step, training throughput and total training time as shown in Table \ref{tab_end2end_performance}. We also report training performance using Branch Parallelism on UniFold and HelixFold. The *-DP method uses 128 accelerator cards only with data parallelism, such as TPUv3 core or A100, while the *-BP method uses 256 accelerator cards in combination with data parallelism and branch parallelism.

Compared with the original AphaFold2-DP, UniFold-DP has been optimized, and the training time has been reduced from 10.961 days to 5.798 days, with an increase of 89\%. We add Branch Parallelism to UniFold to obtain UniFold-BP, which further shortens the training time to 4.181 days, and improves the training performance by 38.67\%.

Similarly, HelixFold has been optimized by operator fusion and tensor fusion to improve the training throughput to obtain the HelixFold-DP performance and the training time has been reduced from 10.961 days to 6.685 days. We use Branch Parallelism to further improve training throughput. The total training time was shortened from 6.685 days to 4.882 days, with an increase of 36.93\%.

Sufficient experimental results show that Branch Parallelism has similar performance improvements in different deep learning frameworks and optimized AlphaFold2 implementations, which is enough to illustrate the effectiveness and generalization of Branch Parallelism.

\section{Conclusion}
%AlphaFold2 is a cutting-edge structure prediction system demonstrating excellent accuracy in CASP14. It draws much attention from academia and industry. 
As the end-to-end training of AlphaFold2 takes lots of computational resources, it is a great burden for the individuals and institutions who are interested in applying AlphaFold2. This paper improves the Evoformer block of AlphaFold2 into Parallel Evoformer, which can be extended to more accelerators by the proposed Branch Parallelism to speed up training. After optimization, the training time was shortened to 4.18 days on UniFold and 4.88 days on HelixFold. We believe that the efficient AlphaFold2 training proposed in this paper can help accelerate research progress in the field of protein structure prediction.

% Use \bibliography{yourbibfile} instead or the References section will not appear in your paper
\clearpage
\bibliography{aaai22}

\end{document}